\documentclass[aps,prb,reprint,superscriptaddress,amsmath]{revtex4-1}
\newcommand{\be}{\begin{equation}}
\newcommand{\ee}{\end{equation}}
\usepackage{graphicx}
\usepackage{dcolumn}
\usepackage{bm}
\usepackage{color}

\begin{document}


\title{Anticorrelations from power-law spectral disorder and conditions for an Anderson transition}


\author{Gregory M. Petersen}
\author{Nancy Sandler}
\email[]{petersengm@gmail.com}
\affiliation{Department of Physics and Astronomy, Nanoscale
   and Quantum Phenomena Institute, and Condensed Matter and Surface Science Program,\\Ohio University, Athens, Ohio
   45701-2979}

\date{\today}

\begin{abstract}

We resolve an apparent contradiction between numeric and analytic results for one-dimensional disordered systems with power-law spectral correlations.  The conflict arises when considering rigorous results that constrain the set of correlation functions yielding metallic states to those with non-zero values in the thermodynamic limit. By analyzing the scaling law for a model correlated disorder that produces a mobility edge, we show that no contradiction exists  as the correlation function exhibits strong anticorrelations in the thermodynamic limit. Moreover, the associated scaling function reveals a size-dependent correlation with a smoothening of disorder amplitudes as the system size increases.

\end{abstract}


\maketitle

\section{Introduction}

The necessary and sufficient conditions for the occurrence of a disordered Anderson transition have received a fair amount of attention in the past half century.\cite{AbrahamsE:1979,izrailev, Kotani_1982,Simon,Kotani,Damanik,Garcia,abrahams201050} Early theoretical work with uncorrelated and short-range correlated disorder potentials identified $\beta = \frac{d\ln{\mathcal{G}}}{d\ln{L}}$ as the relevant scaling function.  If $\beta$ depends solely on the dimensionless conductance $\mathcal{G}$, all the eigenstates of a one-dimensional system will be localized.\cite{AbrahamsE:1979,KramerB:1986} This assumption is known as the single parameter scaling (SPS) hypothesis.\cite{AbrahamsE:1979}  It was realized early on that a rigorous proof of the SPS hypothesis involves an analysis of the full probability distribution of $\ln \mathcal{G}$, or equivalently, its cumulants.\cite{Anderson1} For one-dimensional systems (with uncorrelated disorder potentials), an analysis of these quantities helped determine a relation between the average dimensionless conductance and its variance,
$Var(\ln \mathcal{G}) = -2 \langle \ln \mathcal{G} \rangle$, valid when the SPS hypothesis holds.\cite{Anderson1,Deych_2000} This relationship fails, however, for states at the band edge,\cite{Altshuler1} and at the band center\cite{Schomerus1,kravtsov} which in turn signifies the failure of the SPS hypothesis \cite{Deych_2000} at these energies. These violations allow the possibility of extended states but do not guarantee the presence or absence of a delocalization transition (mobility edge). More complications arise in the analysis of the SPS hypothesis for systems with correlated disorder where violations of SPS appear to be correlated with a crossover between two different scaling regimes for the localization length.\cite{CroyA,Russ_alone,Petersen1}

Due to these limitations, different authors have endeavored to determine the necessary conditions for the existence of an Anderson transition from a more rigorous perspective.  A seminal work was carried out by Kotani in his theory of random ergodic operators.\cite{Kotani_1982,Simon} In it, he was able to show that for the existence of an Anderson transition, Gaussian disorder potentials with correlations must be deterministic (the whole disordered system can be described from the behavior of the potential within a small region).\cite{Kotani_1982,Simon,Kotani,Damanik}  Kotani's theorems also imply that a metallic band can not exist if $\Gamma(x)$, the correlation function for the disordered potential, goes to zero as the distance $x \rightarrow \infty$ as a power law or faster.\cite{Kotani} The theorem thus provides the mathematical justification for the absence of a transition in models with uncorrelated and scale-free disorder distributions in one-dimension.\cite{Russ,CroyA,Petersen1}  More recently, a sufficient condition was also proposed\cite{Garcia} for continuum disordered models: to produce a mobility edge, the disorder potential must be $V(x) \in C^\beta$, with $\beta > 1/2$ and $C^\beta$ representing the class of continuous functions that are $\beta$-differentiable.  Along with these theoretical advances, new experimental studies on ultra-cold atom systems have called into attention the conditions for the existence of an Anderson localization transition in the presence of correlations.\cite{Billy,kuhl} 

Several numerical works on 1D systems with correlated potentials have also addressed the conditions for the existence of extended states.  A broad classification distinguishes three groups of models: discrete,\cite{Phillips} quasi-periodic,\cite{Aubry} and long-range spectral correlations.\cite{DeMoura,kuhl2,russ_affine} Remarkably, the latter, characterized by the use of a disorder spectral density $S(k)$ $\sim 1/k^\alpha$, (with $\alpha$ a measure of the range of correlations) has been shown to produce a localization/delocalization transition. Using a tight-binding model in one dimension, a mobility edge is predicted to appear for disorder strengths $W < 4t$ and $\alpha \geq 2$ (with $t$ the hopping energy).\cite{DeMoura} Despite considerable efforts however, the exact phase diagram for this model for $W > 4t$ remains controversial with contradictory evidence\cite{kaya,shima} regarding the region of parameter space where the transition should occur. Furthermore, numerical studies have remarked the smoothening of the disorder amplitude generated in this model as the thermodynamic limit is reached, suggesting the transition to be order/disorder instead of a true Anderson transition.\cite{russ_affine}

In addition to these issues, the authors in Ref.~\onlinecite{Garcia} pointed out an apparent disagreement between Kotani's theorem and the localization/delocalization transition found in the model with $\sim 1/k^\alpha$ correlations.  The argument presented makes use of the analytic Fourier transform of $S(k)$ to find the corresponding correlation function in real space, with the introduction of an  appropriate cutoff in order to deal with the non-analyticity at $k=0$.  After straightforward algebra, it is found that for values of $\alpha < 1$, the correlation function decays as $~1/x^{1-\alpha}$.  For $\alpha > 1$, however, the Fourier transform renders  exponentially decaying correlations\cite{Garcia} as
\begin{equation}
\Gamma(x) \approx e^{-|cx|^b} \quad \text{for} \quad x \gg a
\label{cor}
\end{equation}
where $c$ and $b$ are parameters and $a$ is the short-range cutoff.  Note that the correlation function goes to zero in the thermodynamic limit in both regimes, i.e., $\Gamma(x)$ appears to violate the necessary conditions for the existence of a mobility edge.\cite{Kotani} The authors thus raised the question of the validity of the Anderson transition found numerically in Ref.~\onlinecite{DeMoura}. 

In order to resolve these inconsistancies and provide some insight into the nature of the transition produced by these correlations, we have carried out a study of the properties of the discrete real-space correlation function as presented in Ref.~\onlinecite{DeMoura}. In this work we present a detailed analytic calculation of the scaling expression for the discrete form of the correlation function, and support the resulting expressions by the numerical evaluation of the correlation function as defined in the original work.
Our findings settle the apparent contradiction between the necessary conditions as imposed by Kotani's results and those obtained in numerical studies of disordered models with $S(k) \sim 1/k^\alpha$ spectral functions. We show that the model contains anticorrelations between the energies of the two most distant sites in the thermodynamic limit. We also confirm that the correlation function obtained in real space is system-size dependent and exhibits an unphysical smoothening of disorder amplitudes as the thermodynamic limit is reached. We argue that these features, some already identified in previous works,\cite{russ_affine} appear to play a significant role in the origin of extended states.

\section{Scaling form of the Correlation Function}

Let us review the procedure to numerically generate random on-site correlated variables for tight-binding Anderson Hamiltonians. To set up notation, we introduce the one-dimensional tight-binding model Hamiltonian with N sites as:
\begin{equation}
H = -t\sum_{<i,j>}^{}c_{i}^{\dagger}c^{}_{j}+h.c. + \sum_{i=1}^{N}\epsilon_{i}c_{i}^{\dagger}c^{}_{i}
\end{equation}
where $t$ is the hopping integral, $c_{i}^{\dagger}$ ($c^{}_{i}$) is the creation (destruction) operator for one fermion at site $i$ and $\epsilon_{i}$ is the on-site disorder energy. $<\dots>$ stands for sum over nearest neighbors sites only. In order to generate the discrete random variables $\epsilon_{i}$, we follow a standard procedure, as presented in Ref.~\onlinecite{osborne}, which is based on colored noise models with a power spectrum $P(\omega_k)=\omega_k^{-\alpha}$. Here $\omega_k = k\Delta \omega$ is a discrete series of frequencies labelled by the integer $k$, $\Delta \omega = 2\pi/T$, $T$ being the total time interval in which the random function is calculated as a discrete time series.\cite{greis,osborne} Within this approach, 
a random variable $X(t_j)$ ($t_j$ is time and $j$ an integer), is defined by the discrete Fourier transform
\begin{equation}
X(t_j) = \sum_{k=1}^{N/2}\Big[a_k \cos(\omega_k t_j) + b_k \sin(\omega_k t_j)\Big]
\end{equation}
where $N=T/\Delta t$,  $\Delta t = (t_{j+1}-t_j)$, and the Fourier coefficients are defined by the power spectral function as: $a_k=\sqrt{P(\omega_k) \Delta\omega}\cos(\phi_k)$ and $b_k=-\sqrt{P(\omega_k) \Delta\omega}\sin(\phi_k)$. Notice that $\phi_k$ is a random independent variable chosen in the interval $[0, 2\pi]$.  Replacing the expressions for the Fourier coefficients above yields:
\begin{equation}
X(t_j) = \sum_{k=1}^{N/2} \sqrt{\omega_k^{-\alpha} \Delta\omega} \cos(\omega_k t_j + \phi_k).
\end{equation}
A map to a tight-binding model in real space is straightforward using the transformations:  $x_j \rightarrow t_j$ and reciprocal variable  $\kappa_{k} \rightarrow \omega_k$
\begin{equation}
X(x_j) = \sum_{k=1}^{N/2} \sqrt{\kappa_{k}^{-\alpha} \Delta\kappa_{k}} \cos(\kappa_{k} x_j + \phi_k).
\end{equation}
In this case, $N$, the numbers of total time-steps is replaced by the total number of lattice sites $N=L/a$ with $L$ the system size, $a$ the lattice constant, $\kappa_{k} = 2\pi k/L$, $\Delta \kappa = 2\pi/L$, and $x_j = ja$.  Replacing these values into the above equation we arrive at 
\begin{equation}
\epsilon_j = \sum_{k=1}^{N/2}\Big|\frac{2\pi}{L}\Big|^{(1-\alpha)/2} k^{-\alpha/2}\cos\Big(\frac{2\pi j k}{N}+\phi_k\Big)
\end{equation}
where we have used $\epsilon_j$ to emphasize that this is the on-site disorder energy (random variable) for the Anderson tight-binding model.  The disorder average of this quantity is given by

\begin{equation}
\langle \epsilon_j \rangle = \sum_{k=1}^{N/2}\Big|\frac{2\pi}{N}\Big|^{(1-\alpha)/2} k^{-\alpha/2}\langle \cos\Big(\frac{2\pi j k}{N}+\phi_k\Big)\rangle
\end{equation}
\begin{equation}
\begin{split}
= \sum_{k=1}^{N/2} & \Big|\frac{2\pi}{N}\Big|^{(1-\alpha)/2} k^{-\alpha/2} \times
\\ & \times \Big(\langle\cos\phi_k\rangle\cos\frac{2\pi j k}{N}-\langle\sin\phi_k\rangle\sin\frac{2\pi j k}{N}\Big).
\end{split}
\end{equation}
The expectation values $\langle \cos\phi_k \rangle = \langle \sin\phi_k \rangle = 0$ resulting in $\langle \epsilon_j \rangle = 0$.

We can calculate the covariance through similar means
\begin{equation}
\begin{split}
\langle \epsilon_m\epsilon_j \rangle = & \sum_{q=1}^{N/2}\sum_{k=1}^{N/2}\Big|\frac{2\pi}{N}\Big|^{1-\alpha} (qk)^{-\alpha/2}  \times
\\ & \times \langle \cos\Big(\frac{2\pi m k}{N}+\phi_k\Big)\cos\Big(\frac{2\pi j q}{N}+\phi_q\Big)\rangle.
\end{split}
\end{equation}
For $k \neq q$, the expectation value can be factorized as $\phi_k$ and $\phi_q$ are independent variables and the only term contributing to the expression is $k=q$:
\begin{equation}
\begin{split}
 = & \sum_{k=1}^{N/2}\Big|\frac{2\pi}{N}\Big|^{1-\alpha} k^{-\alpha} \times
 \\ & \times \langle \cos\Big(\frac{2\pi m k}{N}+\phi_k\Big)\cos\Big(\frac{2\pi j k}{N}+\phi_k\Big)\rangle
 \end{split}
\end{equation}
which can be simplified to yield a covariance of
\begin{equation}
\langle \epsilon_m\epsilon_j \rangle = \frac{1}{2}\sum_{k=1}^{N/2}\Big|\frac{2\pi}{N}\Big|^{1-\alpha} k^{-\alpha}
\cos\frac{2\pi nk}{N}.
\label{sum}
\end{equation}
This even function depends solely on the difference between positions $n=|m-j|$.  The normalized correlation, or auto-correlation, function is defined as (for zero mean)
\begin{equation}
\Gamma(m,j) = \frac{\langle \epsilon_m\epsilon_j \rangle}{\langle \epsilon_j^2 \rangle}
\label{acfdef}
\end{equation}
which when combined with Eq. (\ref{sum}) gives
\begin{equation}
\Gamma(\alpha,n) = \frac{\sum_{k=1}^{N/2}k^{-\alpha}\cos\frac{2\pi n k}{N}}{\sum_{k=1}^{N/2}k^{-\alpha}}.
\label{eq:gamma-n}
\end{equation}
From this point on we impose periodic boundary conditions with the correlation function defined in $n \in [0,N/2]$. This expression has a natural limit of $N \to \infty$ when written in terms of the variable $\gamma=2n/N$ defined in $[0,1]$. The introduction of this variable allows for the explicit removal of the system size dependence in the argument of both sums and emphasizes the scaling form for the correlation function.\cite{Zhang}  

The thermodynamic limit is obtained by taking the upper limit of the sum ($N$) to infinity.
\begin{equation}
\Gamma(\alpha,\gamma) = \frac{\sum_{k=1}^{\infty}k^{-\alpha}\cos \pi \gamma k}{\sum_{k=1}^{\infty}k^{-\alpha}}
\label{final1}
\end{equation}
In order to study this expression, we consider two cases: $\alpha > 1$ and $\alpha < 1$. When $\alpha > 1$  both the numerator and denominator are well defined and Eq. (\ref{final1}) can be rewritten in terms of imaginary exponential functions:

\begin{figure}
\includegraphics[scale=.58]{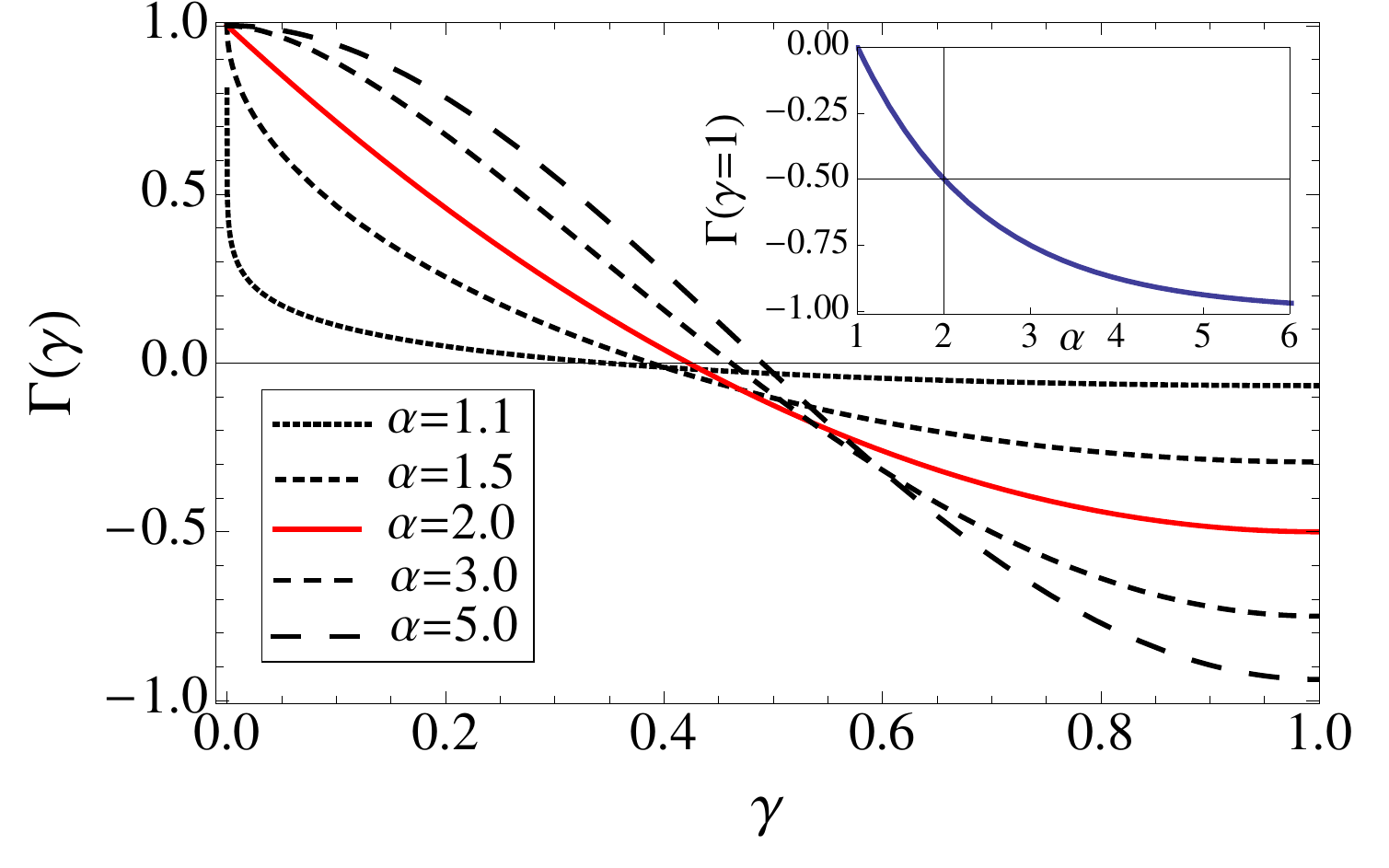}
\caption{Plot of the correlation function as a function of the dimensionless distance $\gamma$.  The red solid line corresponds to the critical value of $\alpha=2$ which is given by Eq. (\ref{aphcrit}).  The other curves correspond to various values of $\alpha$ in Eq. (\ref{final2}). (Inset) Correlation function between the two most distant points ($\gamma=1$).  The origin corresponds to the critical value $\alpha=2$ and yields a correlation of $-1/2$.  For larger values of $\alpha$, the correlation becomes more negative.}
\label{Gvg}
\end{figure}

\begin{equation}
\Gamma(\alpha,\gamma) = \frac{\sum_{k=1}^{\infty}k^{-\alpha}(e^{i\pi \gamma})^k+\sum_{k=1}^{\infty}k^{-\alpha}(e^{-i\pi \gamma})^k}{2\sum_{k=1}^{\infty}k^{-\alpha}}.
\end{equation}
The terms in the numerator are the polylogarithm functions $Li_{\alpha}(z)$ while the denominator corresponds to the Riemann-Zeta sum $\zeta(\alpha)$.  The final expression for $\alpha > 1$ is then
\begin{equation}
\Gamma(\alpha,\gamma) = \frac{Li_{\alpha}(e^{i\pi \gamma})+Li_{\alpha}(e^{-i\pi \gamma})}{2\zeta(\alpha)}.
\label{final2}
\end{equation}

For general values of $\alpha>1$, Eq. (\ref{final2}) cannot be simplified further, however, it takes a very simple expression at $\alpha=2$.  It is important to remark that this value corresponds to the critical value beyond which a band of extended states appears.\cite{DeMoura} By introducing the second order Bernoulli polynomial, $B_2(\gamma/2)$, and using the identity
\begin{equation}
Li_2(e^{i \pi \gamma}) + Li_2(e^{-i \pi \gamma}) = -\frac{(i2\pi)^2}{2}B_2(\gamma/2)
\end{equation}
we obtain
\begin{equation}
\Gamma(\alpha,\gamma) = \pi^2 \frac{B_2(\gamma/2)}{\zeta(2)}
\end{equation}
where $\zeta(2) = \pi^2/6$ and $B_2(\gamma/2) = (\gamma/2)^2 - \gamma/2 + 1/6$.  
Finally, the correlation function reduces to:
\begin{equation}
\Gamma(\alpha=2,\gamma) = \frac{3}{2}\gamma^2 - 3\gamma + 1.
\label{aphcrit}
\end{equation}

This expression is plotted in Fig.\ \ref{Gvg} together with results from Eq. (\ref{final2}) for a few values of $\alpha$ above and below the critical value $\alpha = 2$.  Three features are distinguished in these results:

1.) The function is linear near $\gamma \sim 0$ for the critical value of $\alpha=2$ while the behaviour is convex for $\alpha > 2$ and concave for $\alpha < 2$.  

2.) The correlation function goes negative for $\alpha > 1$. 

3.) The correlation function converges to a non-zero value at the thermodynamic limit of $\gamma=1$ and yields a value of $-1/2$ at the critical value, $\alpha = 2$.

Notice that features 1.) and 2.) occur for smaller values of $\gamma$ and as such can be considered as short-range effects. Feature 3.) however, corresponds to a non-zero value of the correlation function at infinite range. 

\begin{figure}
\includegraphics[scale=.58]{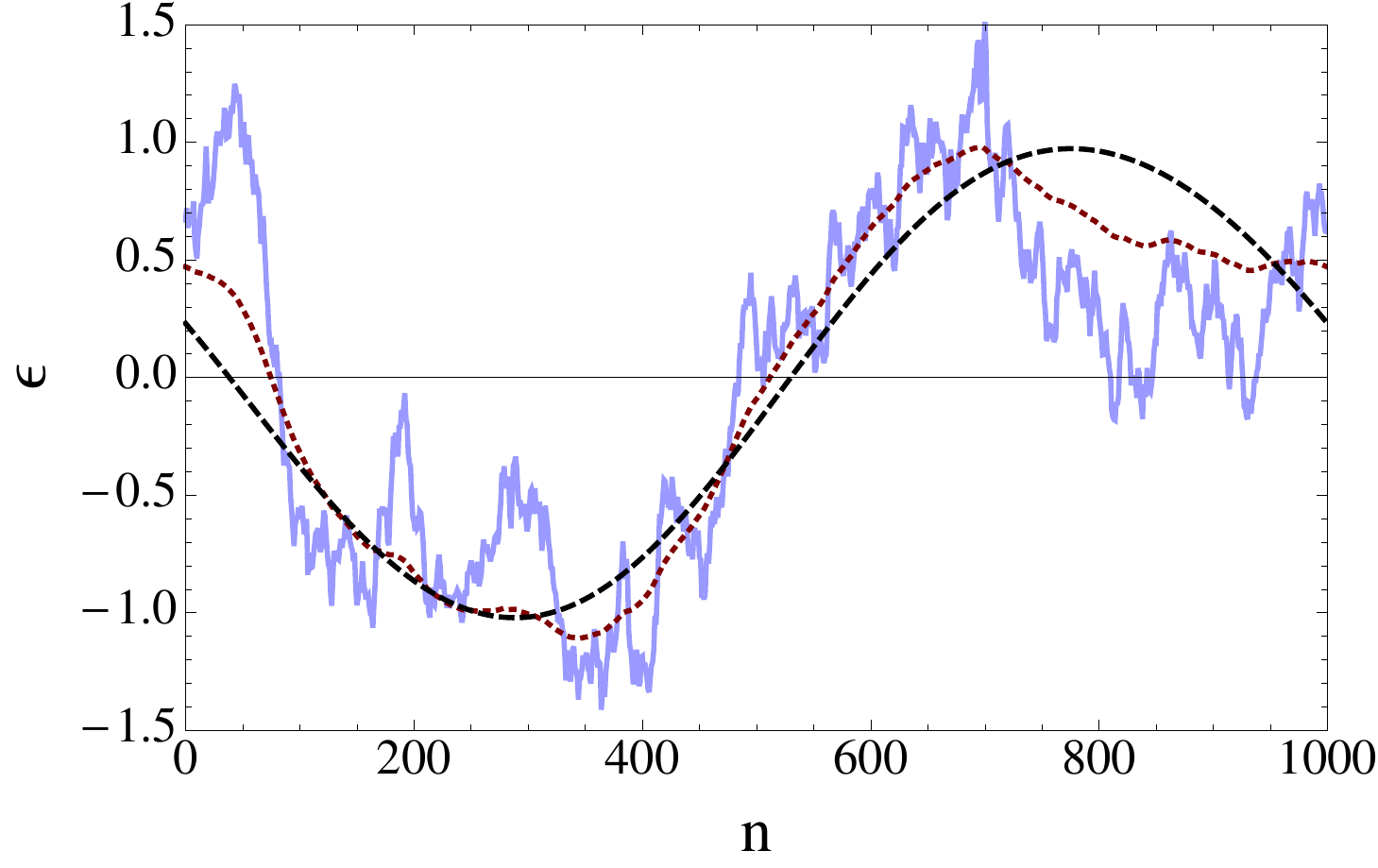}
\caption{Typical disorder realization for $N=1000$, $\alpha=2$ (solid, blue), $\alpha=4$ (dotted, red), and $\alpha=10$ (dashed, black).  Energy is normalized as $\langle \epsilon^2 \rangle = 1$. Plots shows tendency for strong correlations to induce a sinusoidal structure in the disorder configuration.}
\label{Fig3}
\end{figure}

Furthermore, the value of the correlation function between the two most distant points, $\gamma=1$, is
\begin{equation}
\Gamma(\alpha,\gamma =1)= \frac{Li_{\alpha}(-1)}{\zeta(\alpha)}
\end{equation}
where $Li_{\alpha}(-1) = -\eta(\alpha)$ with $\eta(\alpha)$ the Dirichlet-eta function.  We can relate $\eta(\alpha)$ to the Riemann-zeta function through the relationship $\eta(\alpha) = (1-2^{1-\alpha})\zeta(\alpha)$ and obtain
\begin{equation}
\Gamma(\alpha,\gamma =1)= 2^{1-\alpha}-1
\end{equation}
which is plotted in the inset of Fig.\ \ref{Gvg}.

The most remarkable characteristic of these expressions is the existence of a very strong negative correlation between infinitely separated sites.  These negative correlations indicate that any two sites separated by distances of the order of the system size are statistically more likely to have energies with opposite signs.   As a result, an overarching sinusoidal structure begins to develop in the disorder configuration that allows for the appearance of extended low energy states when $\alpha \geq 2$.  We assign this structure to be one of the causes for the emergence of extended states as illustrated in Fig.~\ref{Fig3} which highlight the smoothening of potential amplitudes with increased system size, already proposed in previous works.\cite{Russ} We should note also that a second important element appears to be the value of the correlation strength at infinite distances: our results indicate that it should exceed the value of $\Gamma(\gamma=1)=-1/2$ to produce extended states.  

We turn our focus to the case when $\alpha < 1$.  We return to Eq. (\ref{final1}) and examine the convergence of the numerator and denominator.  The denominator is easily seen to be divergent as the series converges more slowly than the harmonic series for any values of $\alpha \le 1$.  The convergence of the numerator can be seen by applying the Dirichlet convergence test for all $\alpha > 0$. Note that  the series for $\alpha=0$ is not convergent but bounded.  Thus, for all values of $\gamma > 0$ the correlation function is 0 since the numerator is bounded (or convergent) and the denominator diverges.  At $\gamma=0$, the numerator and denominator are equal and the correlation function gives the value 1.  The correlation function for $\alpha \le 1$ can be summarized by:
\begin{equation}
\label{delta}
\Gamma(\alpha,\gamma)=\delta_{\gamma,0}
\end{equation} 
where $\delta_{\gamma,0}$ is the Kronecker-delta function. Thus for values of $\alpha < 1$, the correlation function corresponds to an effective short-range correlated function that precludes the existence of extended states.

Lastly, an additional assumption necessary for the application of Kotani's results is the fact that the random variables generated with correlations given by Eq. (\ref{eq:gamma-n}) must be  Gaussian distributed.  A straightforward test for this condition consists of calculating the fourth cumulant of the distribution to check if its value is zero.  By following a similar procedure as the one outlined above, one can show that the fourth cumulant of the distribution is $K_4 = \frac{3}{4}\zeta(\alpha)^2-\frac{9}{8}\zeta(2\alpha)$ when $\alpha > 1$.  Note that for $\alpha < 1$ the correlations correspond to the regime where no transition takes place. Because it has a non-zero value,  the distribution generated is effectively non-Gaussian thus ruling out the applicability of Kotani's theorems. 

Finally, and in addition to the analysis of the scaling form, it is important to mention that our study of the $\beta$-function and its variance for this model reveals strong violations of the SPS relation $Var(\ln \mathcal{G}) = -2 \langle \ln \mathcal{G} \rangle$ for all energies when $\alpha > 1$. This result further highlights the peculiar nature of the transition found for the set of values $\alpha \geq 2$. 

\section{Discussion}

An identification of the salient features of potential models with  $\sim 1/\kappa^\alpha$ correlations as proposed in Ref.~\onlinecite{DeMoura}, shows that the disordered potential thus generated does not contradict the rigorous conditions put forward in Kotani's work. The scaling form of the correlation function does not vanish in the thermodynamic limit (a necessary condition for the absence of metallic states), but reaches negative values. Moreover, the random energies are not Gaussian variables as required by the theorem. As a consequence, Kotani's theorem can not be used to rule out the existence of a mobility edge for this type of model. Notice that a key result of our analytic derivation in the limit $N \to \infty$ is the scaling form for the correlation function. This scaling emphasizes the fact that the correlation between any two given points is size-dependent, producing a smoothening of disorder amplitudes as the system size increases. This feature, that has been already pointed out in previous works\cite{Russ,Zhang} favors the rise of an ordered potential in the thermodynamic limit, with its corresponding extended-like states.

More insight into the effect of long-range negative correlations can be gained by analyzing the correlation function of the random dimer model\cite{Phillips} with site energies $\epsilon_1, \epsilon_2$. As it is well known, a discrete level with extended states appears  at $\epsilon_1=t$ and $\epsilon_2=-t$. In this case, a numerical evaluation of the correlation between any two points yields $\sim 0.3$. Once again, this value is different from zero and consistent with Kotani's theory.

Finally, these results suggest  that two key ingredients are necessary for the existence of a band of extended states: the crossover to negative correlations in the thermodynamic limit, and a minimum correlation strength between the two most distant points. When considered together, however, they indicate the transition to be an order/disorder one instead of a classical Anderson transition.

\section*{Acknowledgments}

We would like to thank F. A. B. F. de Moura and S. Russ for many insightful discussions.  This work was supported by NSF PIRE and MWN/CIAM grants. We acknowledge the hospitality of the Dahlem Center for Complex Quantum Systems, FU Berlin, where part of this work was completed.

\bibliography{MyBib}

\end{document}